\documentclass[twocolumn,aps,showpacs,floatfix,amsmath,amssymb,prl]{revtex4}

\usepackage{graphicx}

\newcommand{\bc}{\begin{center}}
\newcommand{\ec}{\end{center}}
\newcommand{\be}{\begin{equation}}
\newcommand{\ee}{\end{equation}}
\newcommand{\ba}{\begin{array}}
\newcommand{\ea}{\end{array}}
\newcommand{\beq}{\begin{eqnarray}}
\newcommand{\eeq}{\end{eqnarray}}

\begin{document}

\title{Entanglement entropy at infinite randomness fixed points in higher dimensions}

\author{
Yu-Cheng~Lin$^1$, Ferenc Igl\'oi$^{2,3}$ and Heiko Rieger$^1$}
\affiliation{
$^1$Theoretische Physik, Universit\"at des Saarlandes, 66041 Saarbr\"ucken, Germany\\
$^2$Research Institute for Solid State Physics and Optics, H-1525 Budapest, Hungary\\
$^3$Institute of Theoretical Physics, Szeged University, H-6720 Szeged, Hungary}

\date{\today}

\begin{abstract}
The entanglement entropy of the two-dimensional random transverse
Ising model is studied with a numerical implementation of the strong
disorder renormalization group. The asymptotic behavior of the
entropy per surface area diverges at, and only at, the quantum phase
transition that is governed by an infinite randomness fixed point.
Here we identify a {\it double-logarithmic} multiplicative correction
to the area law for the entanglement entropy. This contrasts with the
pure area law valid at the infinite randomness fixed point in the
diluted transverse Ising model in higher dimensions.
\end{abstract}

\pacs{Valid PACS appear here}

\maketitle

Extensive studies have been devoted recently to understand ground state entanglement
in quantum many-body systems \cite{REV}. In particular, the behavior of various entanglement measures
at/near quantum phase transitions has been of special interest. One of the widely used entanglement 
measures is the von Neumann entropy, which quantifies entanglement of a pure quantum state in 
a bipartite system. Critical ground states in one dimension (1D) are known to have entanglement
entropy that diverges logarithmically in the subsystem size with a universal coefficient 
determined by the central charge of the associated conformal field theory \cite{CARDY}.
Away from the critical point, the entanglement entropy saturates to a finite value,
which is related to the finite correlation length.

In higher dimensions, the scaling behavior of the entanglement entropy
is far less clear.  A standard expectation is that non-critical
entanglement entropy scales as the area of the boundary between the
subsystems, known as the "area law" \cite{AREA,HARMONIC}. This
area-relationship is known to be violated for gapless fermionic
systems \cite{FREE-F} in which a logarithmic multiplicative correction is
found.  One might suspect that whether the area law holds or not
depends on whether the correlation length is finite or
diverges. However, it has turned out that the situation is more
complex: numerical findings \cite{BOSON-NUM} and a recent analytical
study \cite{BOSON-ANA} have shown that the area law holds even for
critical bosonic systems, despite a divergent correlation length.
This indicates that the length scale associated with entanglement may
differ from the correlation length. Another ongoing research activity
for entanglement in higher spatial dimensions is to understand
topological contributions to the entanglement entropy \cite{TOPOLOGY}.
 
The nature of quantum phase transitions with quenched randomness is in
many systems quite different from the pure case. For instance, in a
class of systems the critical behavior is governed by a so-called
infinite-randomness fixed point (IRFP), at which the energy scale
$\epsilon$ and the length scale $L$ are related as: $\ln \epsilon \sim
L^\psi$ with $0<\psi<1$. In these systems the off-critical regions are
also gapless and the excitation energies in these so-called
Griffiths phases scale as $\epsilon \sim L^{-z}$ with a
nonuniversal dynamical exponent $z<\infty$. Even so, certain random
critical points in 1D are shown to have logarithmic divergences of
entanglement entropy with universal coefficients, as in the pure case;
these include infinite-randomness fixed points in the random-singlet
universality class \cite{REFAEL, REFAEL2, SANTACHIARA, YANG, SAGUIA}
and a class of aperiodic singlet phases \cite{APERIOD}.

In this paper we consider the random quantum Ising model in two
dimensions (2D), and examine the disorder-averaged entanglement
entropy. The critical behavior of this system is governed by an IRFP
\cite{QMC,TIM2D} implying that the disorder strength grows without limit
as the system is coarse grained in the renormalization group (RG)
sense. In our study, the ground state of the system and the
entanglement entropy are numerically calculated using a
strong-disorder RG method \cite{FISHER,REV-RG}, which yields
asymptotically exact results at an IRFP. To our knowledge this is the
first study of entanglement in higher dimensional interacting quantum
systems with disorder.

The random transverse Ising model is defined by the Hamiltonian 
\be
H =
-\sum_{\langle i,j \rangle} J_{ij}\sigma_i^z \sigma_{j}^z-\sum_i h_i \sigma_i^x.
\label{eq:H}
\ee
Here the $\{\sigma_i^\alpha\}$ are spin-1/2 Pauli matrices at site $i$ of an $L\times L$
square lattice with periodic boundary conditions. The nearest neighbor bonds $J_{ij}(\ge 0)$
are independent random variables, while the transverse fields $h_i(\ge 0)$ are random or constant. 
For a given realization of randomness we consider a square block $A$ of linear size $\ell$,
and calculate the entanglement between $A$ and the rest of the system $B$, which is 
quantified by the von Neumann entropy of the reduced density matrix for either
subsystems: 
\be
 S=-\textrm{Tr} (\rho_A \log_2 \rho_A) = -\textrm{Tr} (\rho_B \log_2 \rho_B).
\ee


\begin{figure}
{\par\centering \resizebox*{0.4\textwidth}{!}{\includegraphics{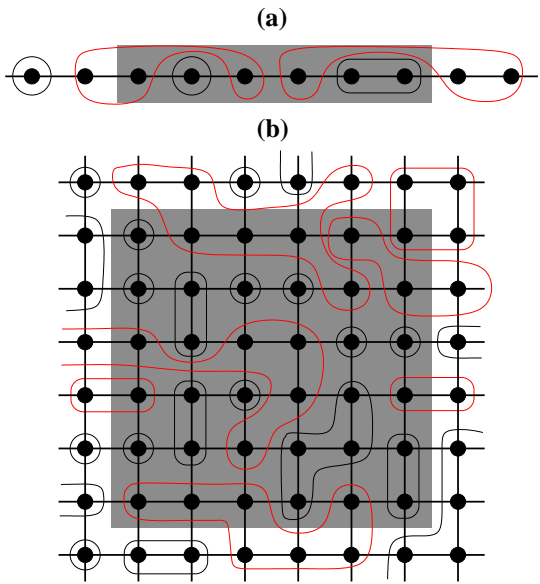}} \par}
\caption{ 
 \label{fig:cluster}
  (color online).
  An example of 
  typical ground state in the random quantum Ising model (a) in 1D, and (b) in 2D;
  it contains a collection of spin clusters of various sizes, which are formed and decimated 
  during the RG. The entanglement of a block (shaded area) is give by the number of 
  decimated clusters (indicated by red loops) that connect the block with the rest of the system. 
 }
\end{figure}


The basic idea of the strong disorder RG (SDRG) approach is as follows
\cite{FISHER,REV-RG}: The ground state of the system is calculated by
successively eliminating the largest local terms in the Hamiltonian
and by generating a new effective Hamiltonian in the frame of the
perturbation theory. If the strongest bond is $J_{ij}$, the two spins
at $i$ and $j$ are combined into a ferromagnetic cluster with an
effective transverse field $\widetilde{h}_{(ij)}=\frac{h_i
h_j}{J_{ij}}$.  If, on the other hand, the largest term is the field
$h_i$, the spin at $i$ is decimated and an effective bond is generated
between its neighboring sites, say $j$ and $k$, with strength
$\widetilde{J}_{jk}=\frac{J_{ij} J_{ik}}{h_i}$. After decimating all
degrees of freedom, we obtain the ground state of the system,
consisting of a collection of independent ferromagnetic clusters of
various sizes; each cluster of $n$ spins is frozen in an entangled
state of the form:
\be
   \frac{1}{\sqrt{2}} (|\underbrace{\uparrow\uparrow\cdots\uparrow}_{n \textrm{ times}} \rangle + 
                       |\underbrace{\downarrow\downarrow\cdots\downarrow}_{n \textrm{ times}} \rangle).
\ee 
In this representation, the entanglement entropy of a block is given by the number of
clusters that connect sites inside to sites outside the block [Fig.~\ref{fig:cluster}].
We note that correlations between remote sites also contribute to the entropy due
to  long-range effective bonds generated under renormalization.

In 1D  the RG calculation can be carried out analytically and
the disorder-averaged entropy $\overline{S}_\ell$ of a segment of length
$\ell$ has been obtained as $\overline{S}_\ell = \frac{\ln 2}{6} \log_2 \ell$ \cite{REFAEL}.
In higher dimensions the RG method can only be implemented
numerically.  The major complication in this case is that the model is
not self-dual and thus the location of the critical point is not
exactly known. To locate the critical point, we can make use of the
fact that the excitation energy of the system has the scaling behavior
$\ln\epsilon\sim L^\psi$ at criticality, while it follows
$\epsilon\sim L^{-z}$ in the off-critical regions. In the numerical
implementation of the SDRG method, the low-energy
excitations of a given sample can be identified with the effective
transverse field $\tilde{h}_{\infty}$ of the last decimated spin
cluster, or with the effective coupling $\tilde{J}_{\infty}$ of the
last decimated cluster-pair.


\begin{figure}
{\par\centering \resizebox*{0.45\textwidth}{!}{\includegraphics{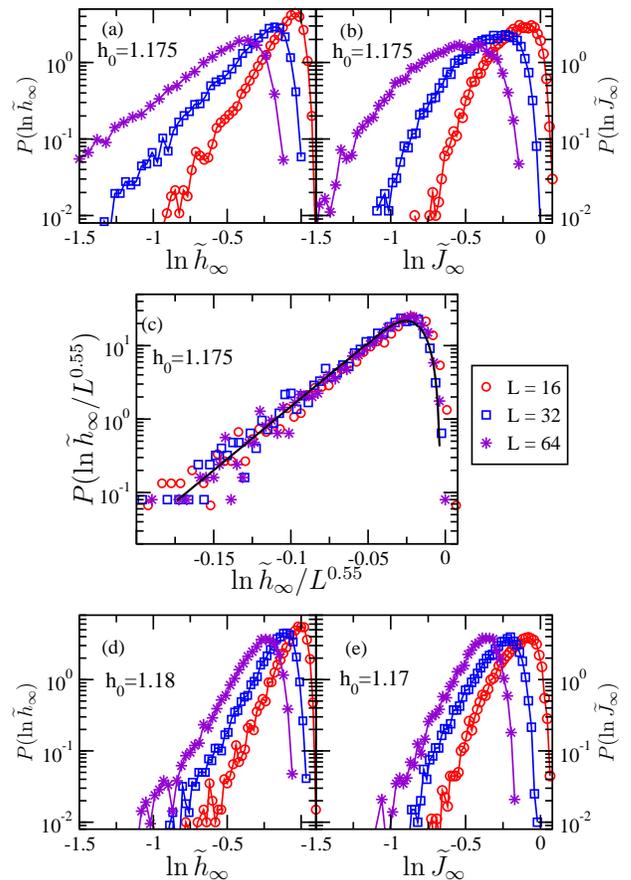}} \par}
\caption{
 \label{fig:gaps}
  (color online).
  The distribution of the last decimated effective log-fields $\ln \widetilde{h}_{\infty}$, 
  and the distribution  of the last decimated effective log-bonds $\ln \widetilde{J}_{\infty}$
  in the RG calculations. 
  At $h_0=1.175$, the distributions, shown in (a) and (b), get broader 
  with increasing system sizes, indicating the RG flow towards infinite randomness, 
  i.e. the system is critical. A scaling plot of the data in (a) 
  using energy-length scaling $\ln \widetilde{h}_{\infty}\sim L^\psi$ with
  $\psi=0.55$ is presented in (c). The solid line is just a guide to the eye.
  The subfigures (d) and (e) show the log-field distribution at $h_0=1.18$ 
  and the log-bond distribution at $h_0=1.17$, respectively;
  the distributions show a power-law decaying tail in the low energy region, 
  which is clear evidence that the system is in the Griffiths phases. 
 }
\end{figure}



\begin{figure}
{\par\centering \resizebox*{0.45\textwidth}{!}{\includegraphics{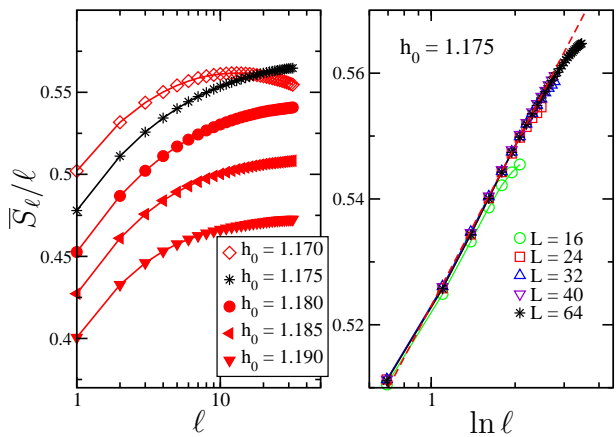}} \par}
\caption{
 \label{fig:S_2D}
  (color online).
  Left panel:
  The disorder averaged block entropy per surface unit $\overline{S}_\ell/\ell$ vs. the linear size of the
  block $\ell$ for a system size $L=64$ for various values of $h_0$.
  We observe that the entropy for $\ell=L/2$ reaches its maximum at the critical point $h_{c}=1.175$ 
  (cf. Fig~\ref{fig:gaps}).
  Right panel:
  The block entropy per surface area vs. $\ln\ell$ on a log-scale for different system sizes 
  $L$ at the critical point. The data show a straight line (guided by the dashed line), 
  corresponding to the scaling obeying the area law with a 
  double-logarithmic correction, as given in Eq.~(\ref{log_log}). 
 }
\end{figure}


In our implementation we set for convenience the transverse fields to
be a constant $h_0$ and the random bond variables were taken from a
rectangular distribution centered at $\overline{J}=1$ with a width
$\Delta=0.5$.  The critical point was approached by varying the single
control parameter $h_0$. Although this initial disorder appears to be
weak, the renormalized field and bond
distributions become extremely broad even on a logarithmic scale
[Fig.~\ref{fig:gaps}] at the critical point $h_0=h_c=1.175$. This
indicates the RG flow towards infinite randomness. Slightly away from
the critical point, both in the disordered Griffiths-phase with
$h_0=1.18$ and in the ordered Griffiths-phase with $h_0=1.17$, the
distributions have a finite width and obey quantum-Griffiths scaling 
$h_\infty\sim L^{-z}$. At the critical point one has IRFP scaling 
$\ln h_\infty\sim L^\psi$ and we estimate the scaling exponent as
$\psi=0.55$, quite close to the value $\psi=0.5$ for the 1D case
\cite{FISHER}.

Now we consider the entanglement entropy near the infinite
randomness critical point. To obtain the disorder-averaged
entanglement entropy $\overline{S}_\ell$ of a square block of size
$\ell$, we averaged the entropies over blocks in different positions
of the whole system for a given disorder realization and then averaged
over a few thousand samples. In Fig.~\ref{fig:S_2D} we show the
entropy per surface unit $\overline{S}_\ell/\ell=\overline{s}_\ell$
for different values of $h_0$. This average entropy density is found
to be saturated outside the critical point, which corresponds to the
area law. At the critical point $\overline{s}_\ell$
increases monotonously with $\ell$, and the numerical data are
consistent with a log-log dependence:
\be
\overline{S}_\ell \sim \ell \log_2  \log_2 \ell
\label{log_log}
\ee
as illustrated in Fig.~\ref{fig:S_2D}. In this way we have identified
an alternative route to locate the infinite randomness critical point:
it is given by the field $h_0$ for which the average block entropy at
$\ell=L/2$ is maximal. Indeed the numerical results in
Fig.~\ref{fig:S_2D} predict the same value of $h_c$ as obtained from
the scaling of the gaps. We note that the same quantity, the position
of the maxima of the average entropy, can be used for the random
quantum Ising chain to locate finite-size transition points
\cite{FSC}.

The log-log size dependence of the average entropy in Eq.(\ref{log_log}) at criticality 
is completely new; it differs from the scaling behavior observed in 2D pure systems,
like the area law, $S_\ell \sim \ell$, for critical  bosonic systems 
\cite{BOSON-ANA,BOSON-NUM}, or a logarithmic multiplicative correction to the area law, 
$S_\ell \sim \ell \log_2 \ell$, as found in free fermions
\cite{FREE-F,BOSON-ANA,BOSON-NUM,F2D-NUM}.  This double-logarithmic
correction can be understood via a SDRG argument: In the
1D case a characteristic length scale $r$ at a given RG step is
identified with the average length of the effective bonds, i.e. the
average size of the effective clusters.  At the scale $r (<\ell)$ the
fraction of the total number of spins, $n_r$, that have not been
decimated is given by $n_r \sim 1/r$ \cite{FISHER}; these active
(i.e., undecimated) spins have a finite probability to form a cluster
across the boundary of the block (a segment $\ell$ in the 1D case)
and thus to give contributions to the entanglement entropy.
Repeating the renormalization until the scale $r \sim \ell$, 
the contributions to the entropy are summed up: 
$\overline{S}_\ell \sim \int^{\ell}_{r_0} {\rm d}r\, n_r  \sim \ln \ell$, leading to
the logarithmic dependence of the 1D model \cite{REFAEL}. 
For the 2D case with the same type of RG transformation with a length
scale $r < \ell$, the fraction of active spins in the renormalized
surface layer of the block is $n_r \sim \ell/r$. Here we have to
consider the situation in which some of these active surface spins
would form clusters within the surface layer and thus contribute zero
entanglement entropy; the number of the active spins that are already
engaged in clusters on the surface at RG scale $r$ is proportional to
$\ln r$, as known from the 1D case, and only $\mathcal{O}(1)$ of the
active surface spins would form clusters connecting the block with the
rest of the system. Consequently, the entropy contribution in 2D can
be estimated as: $\overline{S}_\ell \sim \int^{\ell}_{r_0}{\rm d}r\,
n_r/\ln r \sim \ell \ln \ln \ell$, i.e.\ a double-logarithmic
$\ell$-dependence, as reflected by the numerical data in
Fig.~\ref{fig:S_2D}.

Based on the SDRG argument described above, the
double-logarithmic correction to the area law appears to be applicable
for a broad class of critical points in 2D with infinite
randomness. For instance, the critical points of quantum Ising spin
glasses are believed to belong to the same universality class as
ferromagnets since the frustration becomes irrelevant under RG
transformation, and the same type of cluster formations as observed in
our numerics for the ferromagnet is expected to be generated during
the action of the RG. The entanglement entropy at the IRFP is
completely determined by the cluster geometries occurring during the SDRG.


\begin{figure}
{\par\centering \resizebox*{0.45\textwidth}{!}{\includegraphics{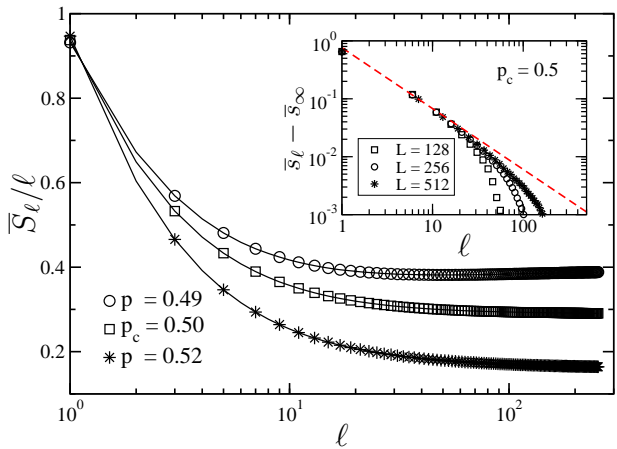}} \par}
\caption{
 \label{fig:dilution} (color online). The entropy per surface area
 $\overline{S}_\ell/\ell=\overline{s}_\ell$ vs. $\ell$ near the percolation
 threshold $p_c=0.5$ for the 2D bond-diluted Ising model at small
 transverse fields for $L=512$. The curves converge to finite values
 for $\ell\to\infty$, corresponding to the area law. The inset shows 
 $\overline{s}_{\ell}-\overline{s}_{\infty}$ as a function of $\ell$.
 $\overline{s}_{\infty}$ is estimated from $\overline{s}_{L/2}$ at $L=512$. 
 The dashed line corresponds to $\ell^{-1}$.
}
\end{figure}


Another type of IRFP in higher dimensions occurs in the bond-diluted
quantum Ising ferromagnet: The Hamiltonian is again given by
(\ref{eq:H}), but now $J_{ij}=0$ with probability $p$ and $J_{ij}=J>0$
with probability $1-p$. At percolation threshold $p=p_c$ there is a
quantum critical line along small nonzero transverse fields, which is
controlled by the classical percolation fixed point, and the energy
scaling across this transition line obeys $\ln\epsilon\sim L^\psi$,
implying an IRFP \cite{TIMDI}. The ground state of the
system is given by a set of ordered clusters in the same geometry as
in the classical percolation model -- only nearest neighboring sites
are combined into a cluster. In this cluster structure, the block
entropy, determined by the number of the clusters connecting the block
and the rest of the system, is bounded by the area of the block,
i.e. $\overline{S}_{\ell}\sim \ell^{d-1}$ with $d$ being
the dimensionality of the system. To examine this, we determined the entanglement
entropy by analyzing the cluster geometry of the bond-diluted
transverse Ising model.  Fig.~\ref{fig:dilution} shows our results for
the square lattice, which follow a pure area-law with an additive
constant: $\overline{S}_\ell=a\ell+b+{\cal O}(1/\ell)$.

To summarize, we have found that the entanglement properties at quantum
phase transitions of disordered systems in dimensions larger than one
can behave quite differently. Generalizing our arguments for the 2D
case, we expect for the random bond transverse Ising systems a
multiplicative $d$-fold logarithmic correction to the area law in $d$
dimensions at the critical point, whereas for diluted Ising model at
small transverse fields the area law will hold in any dimension
$d>1$ at the percolation threshold. Although both critical points are
described by infinite randomness fixed points, the structure of the
strongly coupled clusters in both cases is fundamentally different,
reflecting the different degrees of quantum mechanical entanglement in
the ground state of the two systems. This behavior appears to be in
contrast to one-dimensional systems governed by IRFPs \cite{REFAEL}.

Other disordered quantum systems in higher dimensions might also
display interesting entanglement properties: For instance, the numerical
SDRG has also been applied to higher dimensional random Heisenberg
antiferromagnets which do not display an IRFP \cite{XXX}. The ground states 
involve both singlet spins and clusters with larger moments; therefore, 
we expect the correction to the area law to be weaker than a multiplicative 
logarithm and different from the valence bond entanglement entropy in the
N\'eel Phase \cite{VALENCE}.

Useful discussions with C\'ecile Monthus are gratefully acknowledged.
This work has been supported by the National Office of Research and
Technology under Grant No. ASEP1111, by a German-Hungarian exchange
program (DAAD-M\"OB), by the Hungarian National Research Fund under
grant No OTKA TO48721, K62588, MO45596.

\end{document}